\documentclass[conference]{IEEEtran}
\IEEEoverridecommandlockouts
\usepackage{cite}
\usepackage{amsmath,amssymb,amsfonts}
\usepackage{algorithmic}
\usepackage{graphicx}
\usepackage{textcomp}
\usepackage{xcolor}
\usepackage{svg}
\usepackage{soul}
\usepackage{listings}
\usepackage[colorinlistoftodos,prependcaption,textsize=tiny]{todonotes}
\usepackage{lipsum}                     % Dummytext
\usepackage{xargs}                      % Use more than one optional parameter in a new commands
\usepackage[pdftex,dvipsnames]{xcolor}  % Coloured text etc.
\usepackage{hyperref}
\newcommandx{\note}[2][1=]{\todo[linecolor=blue,backgroundcolor=blue!25,bordercolor=blue,#1]{#2}}

\def\BibTeX{{\rm B\kern-.05em{\sc i\kern-.025em b}\kern-.08em
    T\kern-.1667em\lower.7ex\hbox{E}\kern-.125emX}}
\begin{document}

\title{Industrial Semantics-Aware Digital Twins:\\A Hybrid Graph Matching Approach for\\Asset Administration Shells\\
}

\author{\IEEEauthorblockN{Ariana Metovi\'{c}*, Nicolai Maisch, Samed Ajdinovi\'{c}, \\Armin Lechler, Andreas Wortmann and Oliver Riedel}
\IEEEauthorblockA{Institute for Control Engineering of Machine Tools and Manufacturing Units (ISW),\\
University of Stuttgart, Seidenstr.36, 70174 Stuttgart, Germany\\
\ ariana.metovic@isw.uni-stuttgart.de\\
*Corresponding author}
}

\maketitle

\begin{abstract}
 Although the Asset Administration Shell (AAS) standard provides a structured and machine-readable representation of industrial assets, their semantic comparability remains a major challenge, particularly when different vocabularies and modeling practices are used. 
 Engineering would benefit from retrieving existing AAS models that are similar to the target in order to reuse submodels, parameters, and metadata. 
 In practice, however, heterogeneous vocabularies and divergent modeling conventions hinder automated, content-level comparison across AAS. 
 This paper proposes a hybrid graph matching approach to enable semantics-aware comparison of Digital Twin representations. 
 The method combines rule-based pre-filtering using SPARQL with embedding-based similarity calculation leveraging RDF2vec to capture both structural and semantic relationships between AAS models.
 %\hl{ We discuss possible difficulties for future implementations.}
% \todo{machen wir das, eine evaluation?} We evaluate the approach on multiple heterogeneous AAS datasets to demonstrate its capability to identify similar AAS models and submodels despite vocabulary mismatches. 
 This contribution provides a foundation for enhanced discovery, reuse, and automated configuration in Digital Twin networks.
\end{abstract}


\section{Introduction}

The increasing digitization of industrial production is fundamentally transforming how assets and processes are managed. 
A key enabler of this transformation is the Digital Twin, which establishes a dynamic connection between physical assets and their digital counterparts, enabling monitoring, analysis, and control across the product lifecycle \cite{Soori.2023}.
Various approaches and data models have emerged in order to provide the foundation for digital twins, such as the Asset Administration Shell (AAS) \cite{Zhang.2025}.
The AAS is a standardized metamodel that provides a machine-interpretable description of assets within Industry 4.0 environments, thereby ensuring interoperability across the manufacturing industry and beyond \cite{IDTAWorkstreamSpecificationofAAS}.

In practice, however, AAS models often rely on different vocabularies and modeling conventions. 
Although these models may be syntactically valid, they frequently lack semantic alignment, making automated comparison and reuse difficult. 
%\todo{diesen Satz hier klarer machen, also warum und für weche fälle braucht man das?} Especially when thousands of products have their own digital representation, an interoperable approach for querying and knowledge discovery is needed to identify and then reuse existing models as examples\cite{MohammadHosseinRimaz.2024}.

An industrial use case that often involves significant manual effort and potential duplication arises when modeling a new product. 
Since each product must be represented individually, efficiently identifying and reusing existing models and assets can greatly simplify engineering tasks. 
Reusing available submodels, parameters, and metadata not only reduces engineering effort but also ensures consistency across products. 
However, even with semantic annotations, identifying appropriate AAS models within a large pool of assets can still be challenging.
The resulting heterogeneity increases manual integration efforts for example, when equipment from multiple suppliers must be combined into a unified production system.

In the fields of Semantic Web and Machine Learning, promising technologies such as ontologies and graph-based \cite{Liu.2007} or vector-based representations \cite{Li.2021} offer potential for improving AAS discovery.
This paper proposes a hybrid graph-matching approach that combines rule-based semantic pre-filtering with vector-based similarity computation to enable more accurate matching of heterogeneous AAS models. 
While graph-based discovery (e.g., using query languages such as SPARQL) captures the structural composition of AAS models vector-based methods provide semantic comparability. 
%\todo{Ggf. noch deutlicher machen, warum man das jetzt braucht} 
Because brownfield AAS models often lack uniform structures and vocabularies, combining both perspectives enables flexible and robust discovery across heterogeneous AAS networks and repositories.
This approach not only improves the accuracy of matching relevant AAS instances but also enhances their reusability and interoperability within Digital Twin networks.

%While these techniques address structural and semantic aspects, they are not tailored to the specific requirements of AAS-based Digital Twins. 
%To overcome this limitation, this paper proposes a hybrid graph matching approach that combines rule-based semantic pre-filtering based on graph representations with vector-based computation. 
%The proposed method enables more accurate discovery and matching of heterogeneous AAS models and enhances interoperability through re-use in Digital Twin networks as long as they are based on AAS while reducing engineering effort. 

%\hl{The research gap hereby addresses that although AAS data can already be represented, stored and queried using RDF technologies, no existing approach enables true content-level comparability between different AAS instances. SPARQL supports queries but cannot uncover semantically related AAS expressed with different vocabularies or modeling conventions. Conversely, embedding-based methods capture semantic proximity but do not enforce the structural requirements needed for practical reuse. The core research gap therefore lies in the absence of a unified method that combines structural constraint handling with semantic similarity scoring. Such an approach is useful for reliably identifying relevant and reusable AAS within RDF repositories.}

The remainder of this paper is structured as follows. 
Section II introduces the fundamental concepts and terminology relevant to Digital Twins and semantic modeling. 
Section III reviews the state of the art in semantic matching, as well as graph- and vector-based similarity methods, and highlights the existing research gaps.
Section IV presents the proposed hybrid graph matching approach, followed by the modeling and learning objectives in Section V and the discussion of testing pipeline and weaknesses in Section VI.
Section VII concludes the paper and outlines directions for future work.
\section{Fundamentals}

In this section the formal structure of the AAS and the function of Resource Description Framework (RDF) as a standard model for data interchange on the Web is outlined. 

\subsection{Asset Administration Shell}\label{heading:aas}

\begin{figure}[t!]
    \centering
     \includegraphics[width=0.8\linewidth]{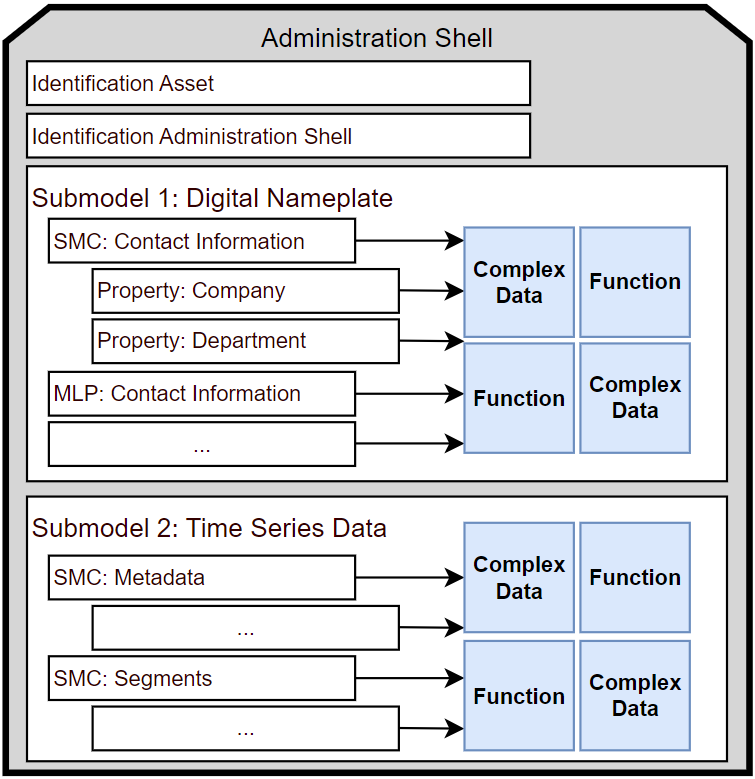}
    \caption{Example of the basic Structure of an Asset Administration Shell based on \cite{IDTAWorkstreamSpecificationofAAS}}
    \label{fig:AasArch}
\end{figure}

The AAS is a metamodel standard for digitally representing industrial assets.
The basic structure of the AAS consists of information about both the asset and the AAS itself (e.g., unique identification details), as well as submodels that contain more detailed information depending on the specific use cases \cite{IDTAWorkstreamSpecificationofAAS}. (Fig. \ref{fig:AasArch}) shows an example AAS featuring a digital nameplate and time series data, with corresponding submodel templates available from the IDTA\footnote{https://industrialdigitaltwin.org/en/content-hub/submodels}.

Serialized in standard formats such as JSON or XML, or in the custom AASX format \cite{IDTAWorkstreamSpecificationofAAS.e}, the AAS can be flexibly distributed, modified, and extended.
Each information element of the AAS (such as name or referenced submodels) and its submodels (such as nameplate or energy usage) is identified using unique identifiers and can be semantically described using an internal concept description or references to standardized vocabularies (e.g., IEC CDD \cite{IEC_CDD} or ECLASS \cite{ECLASS}).
These annotations can be used to discover assets with specific attributes such as finding all pumps on a shop floor or identifying all assets older than five years.

\subsection{Knowledge Graphs and Resource Description Framework}
A knowledge graph is a technique for structuring and linking information through explicit relationships between entities, forming a modular and extensible graph with nodes and edges \cite{HeikoPaulheim.2016}. 
%\todo{???} 
An ontology is an explicit specification of a conceptualization\cite{GRUBER1993199}.
A knowledge graph organizes data through nodes and relationships, while an ontology provides the formal vocabulary and rules that define the meaning of nodes and relations.
% Quelle überprüfen, ob sie immer noch stimmt
The W3C standard RDF is used as a technical implementation for describing and interlinking data in the Semantic Web \cite{RDFWorkingGroup.2014}.
RDF describes information in form of triples that combine to form graphs and enable a flexible, extensible representation \cite{Pan.2017}. 
An RDF statement has the following structure: 
\[
\langle \text{subject} \rangle \quad \langle \text{predicate}\rangle \quad \langle\text{object} \rangle
\]

An RDF statement expresses a relationship between two resources, the subject and the object. The relationship is directed from the subject to the object and the predicate is describing the relationship\cite{RDFWorkingGroup.2014}.
RDF can be serialized in various formats, such as RDF/XML, N-Triples, JSON-LD or Turtle\cite{RDFWorkingGroup.2014b}.
%\todo{eher sagen, dass RDF json-ld based serialisiert ist oder in spezifischen Formaten wie Turtle [Quelle]... "RDF can be serialized in..."}
%RDF offers a variety of syntax notations, one of the widely used notations being the Terse RDF Triple Language (Turtle)\cite{RDFWorkingGroup.2014b}.
%The Web Ontology Language (OWL) extends RDF functionalities by providing formally defined classes, properties, and data values \cite{OWLWorkingGroup.}.
SPARQL is a query language for RDF data, used for querying and manipulating RDF graph content on the Web or in an RDF store. It supports SELECT queries (return variable bindings), ASK queries (e.g.  boolean) and CONSTRUCT queries (construct new RDF graphs from a query result)\cite{SPARQLWorkingGroup.}.
%\todo{noch bisschen info zu sparql}
%RDF-based ontologies are used in a variety of domains already. 
%One area of research is therefore the usage of RDF in combination with AAS within Industry 4.0. An important contribution is article \cite{Rongen.2023}. 
% Das hier bitte ändern. Der Satz (mit Rongen) sagt nichts aus. Dass man RDF mit AAS und I4.0 nutzt, ist für ich aber eher schon SotA

\section{State of The Art}
%One area of research is the usage of RDF in combination with AAS within Industry 4.0\cite{Rongen.2023}.
\subsection{RDF Technology for AAS}\label{heading:conversion}
Besides the serialization as JSON or XML, the AAS may be represented as an RDF graph \cite{Rongen.2023}, enabling semantic reasoning on the AAS metamodel. 
%\cite{Bader.}
%By combining the descriptive AAS metamodel with the reasoning capabilities of RDF, this approach enables semantic discovery functionalities \cite{Rongen.2023}.
\cite{MohammadHosseinRimaz.2024} proposed a Python package for converting AAS data into RDF format\footnote{GitHub of the converter: https://github.com/mhrimaz/py-aas-rdf}. 
It also provides a graphical user interface for editing and showing converted AAS.
However, there remain challenges in the conversion, such as the lack of support for ordered elements in the RDF representation of an AAS \cite{MohammadHosseinRimaz.2024}.
%\note{Stimmt das noch mit deiner Aussage überein oder kann das Tool das abfangen?}
Other AAS tools such as Eclipse BaSyx \cite{EclipseBaSyx.2024}, which supports AAS hosting, or AAS4J \cite{EclipseFoundation.}, which enables AAS-based development in Java, feature RDF only as experimental plugins\footnote{GitHub branch Experimental/Adapter/RDF: https://github.com/eclipse-basyx/basyx-python-sdk/tree/Experimental/Adapter/RDF} or planned extensions\footnote{GitHub Eclipse AAS4J: https://projects.eclipse.org/projects/dt.aas4j}.
The platform metaphactory\footnote{https://metaphacts.com/metaphactory} provides an application layer on top of RDF knowledge graphs that integrate ontologies and standard dictionaries (e.g., IEC CDD, ECLASS) together with AAS. It offers a SPARQL-based discovery stack and a user interface for exploring entities and relations. An integrated AI service enables natural-language questions that are translated into SPARQL, making semantic querying accessible to non-native SPARQL users \cite{MohammadHosseinRimaz.September2025}.

\subsection{RDF Stores and Querying}
Within Semantic Web technologies, extensive RDF stores comprising numerous RDF triples are established\footnote{Including Ontotext GraphDB, OpenLink Virtuoso, Apache Jena Fuseki, Eclipse RDF4J, Stardog or Amazon Neptune}.
This infrastructure supports complex querying, information discovery, and reasoning through SPARQL querying \cite{Sagi.2022}.
The main components of an RDF store are the repository and the middleware to continuously communicating with the underlying repository \cite{Hertel.2009}. 
%\cite{Modoni.2014} found that various implementations of RDF stores are suitable as a backbone of semantic applications that need to store and process large amount of RDF data in a reliable manner and that most of them provide support for important requirements such as data protection, information privacy and security.
Various implementations of RDF stores are suitable as a backbone of semantic applications that need to store and process large amount of RDF data in a reliable
manner and that most of them provide support for important requirements such as data protection, information privacy and security\cite{Modoni.2014}.
%Widely used RDF stores\footnote{Including Ontotext GraphDB, OpenLink Virtuoso, Apache Jena Fuseki, Eclipse RDF4J, Stardog or Amazon Neptune} provide SPARQL query access via HTTP REST endpoints, support standard frameworks such as RDFS and OWL, and often include SHACL-based validation to ensure data quality.

%\cite{Nitta.} proposed architectures for RDF store management software, including indexing strategies. 
%They present a detailed classification applied to major systems (e.g., Virtuoso, RDF-3X, Hexastore, Jena, AllegroGraph) and identify remaining gaps in distributed processing for efficient querying. %\note{Hier sagen, was die besonders gemacht haben oder welche benchmark sie erreicht haben. Nur nennen des Frameworks bringt keinen Mehrwert. Entweder sagen, welche besondere Methodik vorgestellt wurde oder welche Benchmark erreicht wurde}
%\note{Genau so vorgeben für Voigt und Lam}

\cite{MartinVoigt.} and \cite{AnNgocLam.2023} present SPARQL benchmark studies on real-world datasets, demonstrating its scalability and efficiency for querying complex data.

\subsection{Comparability of RDF Graphs and Data}
%In addition to enabling data discovery, RDF stores support the semantic comparability of entities within a dataset. 
By representing information as semantic statements, RDF stores link nodes, allowing them to be placed in relationships.
%In this context there is a wide set in how to compare semantic statements within the set, 
A variety of techniques exist to compare such graphs, ranging from structural matching based on the formal structure of the graph(s) over semantics-aware alignment based on semantic annotations up to embedding-based similarity.% \note{Hier noch ergänzen. Ich hab das mit dem representation learning nicht verstanden. Gerne mit Quelle hinterlegen, dass es verschiedene Sachen gibt.}

%Comparing RDF graphs (and their subgraphs) spans a spectrum from purely structural matching to semantics-aware alignment and representation-learning.
%Classic Graph Edit Distance (GED) measures the minimum cost to transform one (sub)graph into another\cite{Serratosa.2021}.

\cite{HaiJinLiHuangPingpengYuan.2010} proposes a graph-based method for efficient duplicate detection in RDF datasets by comparing subgraphs for similarity and aggregating these to detect duplicates.
While conceptually related, the approach focuses on data cleansing rather than semantic querying or graph-based knowledge retrieval. %No statement was made as to how the approach can be used for these aspects.

%\note{irgendwie kommt es random, dass jetzt auf deine Arbeit verwiesen wird... kann man das besser formulieren im sinne von "Es wurde keine Aussage getroffen, wie der Approach für semantic querying oder graph-based knowledge retreival genutzt werden kann"?}
%The authors of \cite{HaiJinLiHuangPingpengYuan.2010} propose a approach called K-Radius Subgraph Comparison (KSC), a graph-based method for efficient duplicate detection in RDF datasets. Their approach models RDF data as hierarchical graphs and compares pairwise similarity with context comparison over k-radius subgraphs, and merges these into a total similarity to decide duplication. However, KSC targets data cleansing and duplicate detection, whereas the focus here lies on semantic querying and graph similarity for knowledge retrieval, making it a conceptually related but functionally distinct approach.

\cite{WeiguoZheng.} utilizes Graph Edit Distance (GED), a method that measures the minimum cost to transform one (sub)graph into another\cite{Serratosa.2021}, and enriches the costs with predicate/type similarity to reflect RDF labels, therefore combining structural and semantic similarities.% and inherit GED’s exponential worst-case complexity, therefore combining structural and semantic similarities.
%\note{wenn wir das vorher in verhältnis zur arbeit setzen, dann hier am besten auch KURZ.}
%In the literature \cite{WeiguoZheng.} a RDF-tailored variant is proposed, that enriches edit costs with predicate/type similarity to reflect RDF labels und inherit GED’s exponential worst-case complexity. Incorporating both structural and semantic similarities, the novel similarity measure \textit{}{semantic graph edit distance} (SGED) is proposed.

%Ontology matching methods like LogMap\cite{ErnestoJimenezRuizandBernardoCuencaGrau.}, PARIS\cite{FabianM.SuchanekSergeAbiteboulPierreSenellart.}, RiMOM\cite{Li.2009} are mentioned here because they are suited to reconcile heterogenous vocabularies and can be used as an optional alignment pre-step. However, since they do not provide a generic subgraph similarity score, they are not of primary use within the objectve.
%\note{Wenn die Sachen nicht wichtig sind, dann lass sie weg.}

\cite{10.1007/978-3-319-68288-4_31} introduces a formal framework for pairwise entity comparison in RDF graphs, generating explicit similarity or difference queries instead of producing a global similarity score suitable for large-scale retrieval.
%\cite{10.1007/978-3-319-68288-4_31} introduces a formal framework for entity comparison in RDF graphs, addressing the task of identifying similarities and differences between entities beyond standard SPARQL querying. The framework operates pairwise and generates the explicit similarities or differences in form of queries instead of a single global similarity score over large graphs.
%\note{versteh ich irgendwie nicht}
% They provide fine-grained, explainable comparisons but operate pairwise and do not yield a 

\subsection{Vectorial Matching Approaches}
\begin{figure}
    \centering
    \includegraphics[width=1\linewidth]{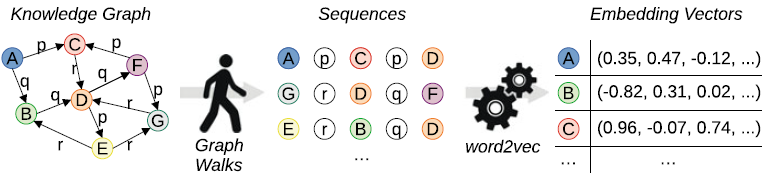}
    \caption{Overall workflow of RDF2vec\cite{Paulheim.2023}}
    \label{fig:rdf2vec}
\end{figure}

With the rise of machine learning, particularly language models, the embedding-based approach RDF2vec \cite{Paulheim.2023} was introduced. 
It generates vector representations from random walks or triple sequences, such that semantic and structural proximity in the RDF graph is reflected as geometric closeness in the embedding space (Fig.\ref{fig:rdf2vec}).
In RDF2vec, extracted sequences of entities and relations from the graph are then fed into the Word2vec algorithm, producing embedding vectors for all entities encountered in these walks. 
Graph similarity is then computed using distances between these vectors.
%\note{RDF2vec or Word2Vec? ich bin verwirrt?} 

In the field of Industry 4.0, there are ongoing efforts to semantically match information contained within AAS, or even across different data models, using natural language processing techniques to achieve semantic interoperability \cite{Beermann.2023}. 
By generating vector representations of AAS and their data (embeddings), semantic similarity can be quantified \cite{Xia.2022}. However, this approach primarily handles textual representations of AAS, which can pose limitations when analyzing large numbers of instances.
\cite{Ajdinovic.icmimt.2025} presented a Retrieval-Augmented Generation (RAG) \cite{RAG_2020}–based semantic matching algorithm that bridges multiple industrial information models. By utilizing vector representations of instances, the study demonstrated how language models can identify semantic similarities to automatically integrate heterogeneous data entities.

%First the sequences of entities and relations are extracted from the grapha and the sequences are fed into the word2vec algorithm to create embedding vectors for all entities in the knowledge graph\cite{Paulheim.2023}. After the method, the graph similarity is measured with vector distances. 

\subsection{Research Gaps}
Although RDF has been used for AAS discovery, methods for automatically comparing AAS across large RDF stores are still lacking.
%Semantic Web techniques enable querying and semantic comparison. %, and when combined with machine-learning-based embeddings, even deeper semantic reasoning becomes possible. 
Current systems expose rich metadata but do not support automated content-level comparisons. 
SPARQL queries return explicit structural matches but cannot reveal similarities between conceptually related entities expressed with different vocabularies or naming conventions. 
Semantic matching approaches based on embeddings perform well for similarity-based matching tasks and are able to capture only local graph structure
but may have computational limitations in large, complex data stores and are restricted to textual information.
By combining advantages of both technologies semantic matching and discovery can be improved. 
In summary, RDF-based discovery can be enhanced with vector-based representations to leverage semantic features for more effective AAS discovery.
We therefore propose a hybrid method to quantify similarity between AAS graphs/subgraphs and automatically retrieve comparable AAS models.
This integrates

\begin{enumerate}
    \item the structural relationships of the RDF graph,
    \item the semantic meanings captured by ontologies, and 
    \item synonymy and terminological variation, 
\end{enumerate}

thereby providing a robust foundation for content-level comparability of AAS in RDF.

%\note{Ich ziehe den Use Case Teil hoch in die Intro}

%\input{src/03_Motivation}

\section{Architecture}

\begin{figure}
    \centering
    \includegraphics[width=0.95\linewidth]{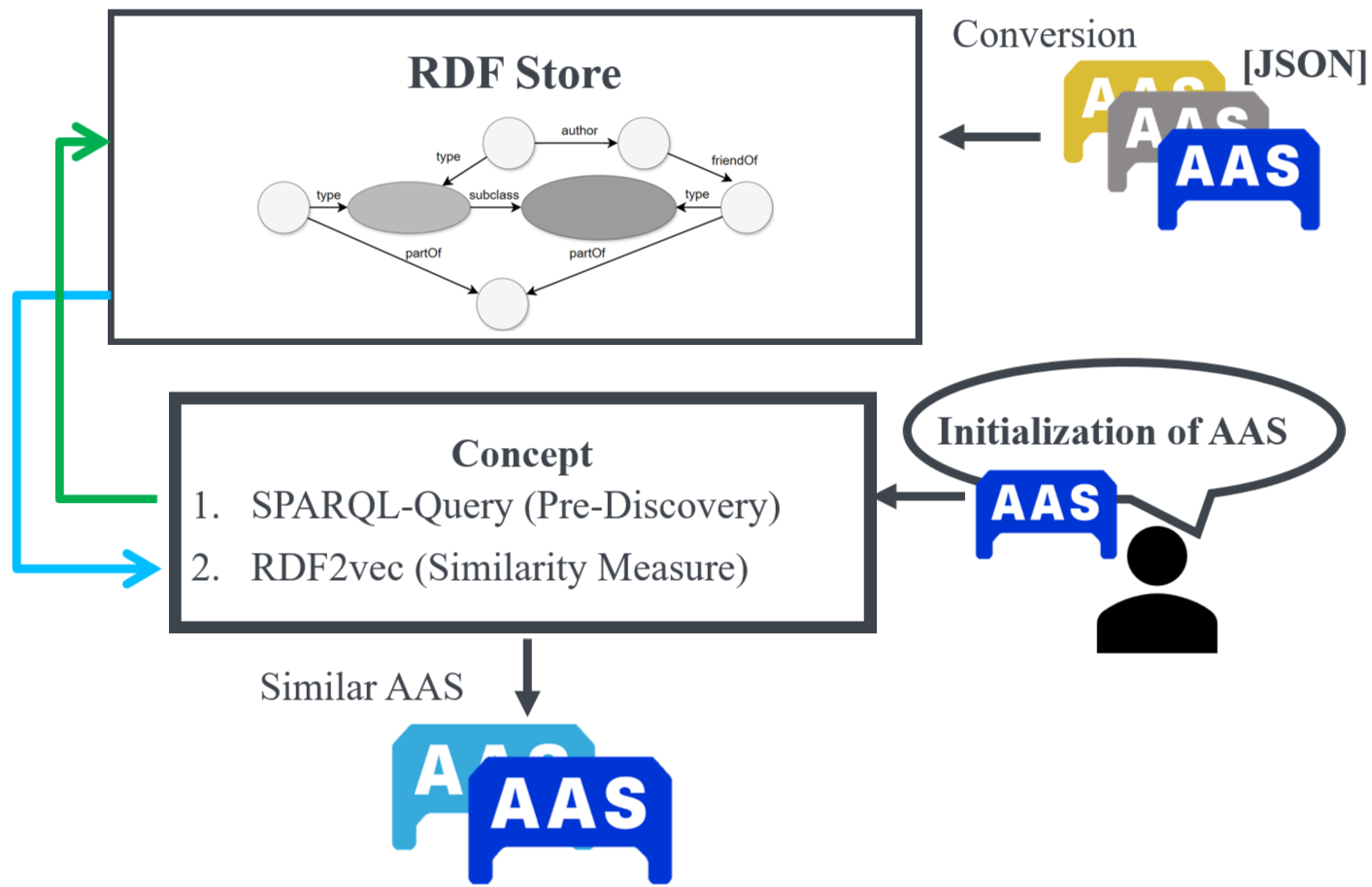}
    \caption{Concept of the the proposed Hybrid Graph Matching Approach}
    \label{fig:concept}
\end{figure}

%To address these challenges, we propose a hybrid graph matching approach for the discovery and retrieval of AAS instances. 
The method combines pre-filtering using SPARQL over RDF with vector-space similarity via RDF2vec, enabling identification of AAS models that are semantically similar to a target while leveraging the strengths of both paradigms (Fig. \ref{fig:concept}). 
The objective is to enable constraint handling and context-sensitive semantics.

While RDF’s structured querying enables navigation through the AAS composition, vector-based semantic comparisons can resolve similar data such as "power input" and "electric power". % in the digital twins of electric motors.
%Also, \cite{miny2025} ... \todo{hier beschreiben, was Miny sagt}
%\hl{
%Mentioned in this article are structural and semantical similarities. 
%The general structure of a AAS in explained in Chapter} \ref{heading:aas}. \hl{A structural comparison of two AAS for this approach is looking at their composition. As an example, standardized submodel templates define the internal structure of a submodel, }which ensures a similar structure for submodel instances\cite{Zhang.2025}.
%\hl{A semantic comparison is looking at the meaning of word or entities within the graph. For instance, semantically equal representations of the integer 2 are "2", "+2" or "0002". The order of submodels and submodel elements does not change the semantical meaning either, which is why differently ordered AAS with the same content should semantically still be the} same. \cite{miny2025}
%\todo{würde den satz weglassen}Therefore, we define similar AAS instances as those that share comparable structures and semantic characteristics. 
In terms of AAS modeling, two AAS instances may both implement the same submodels, but additionally describe related properties possibly using different labels. 
Because they have a similar set of submodels and their contexts are semantically aligned, the query algorithm will most likely match.
%electric motors
%Digital Nameplate and Timeseries templates
%such as torque, rotational speed, or voltage, 
%This similarity provides the basis for transferring metadata, parameter sets, or entire submodels into the user’s incomplete AAS.

%\hl{the comparison of the included submodels and elements.  --> Vllt hier Mengenlehre anwenden??!!}
%\hl{semantical comparison: --> Paper von Miny zitieren mit Bsp der Nummer 2!!!}

%\subsection{Pre-discovery via SPARQL}
\subsection{Data Storage and pre-discovery via SPARQL}

Existing AAS instances and templates are provided in JSON and converted into RDF using the mapping tool \cite{MohammadHosseinRimaz.2024}. The resulting graphs are persisted in a data store and exposed via SPARQL endpoints, yielding a queryable representation of all AAS and enabling pre-discovery as the concept's first step.

%\todo{das wirkt auf mich wie eine dopplung von oben (IV Anfang)}
In industrial scenarios, the AAS used as input for a workflow is incomplete: it may lack a specific submodel (e.g. Timeseries) or standardized parameters (e.g. a torque parameter defined according to ECLASS or IEC CDD), while still containing essential information about the asset (e.g. Digital Nameplate). The goal is not to compare this partial AAS with all available AAS, but only with candidates structurally fulfilling missing requirements. For this reason, SPARQL is used as an explicit constraint mechanism rather than a similarity measure. Missing information is expressed as SPARQL constraints (e.g. required submodels, properties). This yields a filtered AAS set, reducing computation and avoiding irrelevant matches while ensuring only structurally valid candidates proceed to semantic scoring.
%All further semantic knowledge is handled by the embedding-based step. 

%\hl{Because of the metamodel and available IDTA templates, the SPARQL query expresses only the required structure of the missing part—for example, “the AAS must contain the Timeseries submodel” or “must include a torque parameter defined according to known vocabularies like ECLASS or IEC CDD”. All further semantic knowledge is handled by the embedding-based step. Thus, the SPARQL stage does not attempt to identify semantically equivalent concepts, nor does it enforce structural similarity between entire AAS graphs. Instead, it removes AAS that cannot possibly satisfy the user’s requirements. This reduces computation and avoids irrelevant matches while ensuring that only structurally valid candidates proceed to semantic scoring.}

%\todo{"During Engineering process ..."}
For instance, during the engineering process a user's AAS is missing the Timeseries submodel. The user wants to find existing AAS instances that include this submodel so that missing parameters can be reused. Listing \ref{lst:query} presents a pseudo-SPARQL query that implements this filtering pattern and returns only AAS containing the required submodel. The result is a set of plausible candidates that are passed to the next stage.

%To illustrate the aspect, we are using an example of a search for an electric motor with time series data within the RDF store.

%We first restrict candidates to AAS that include a Digital Nameplate submodel, as electric motors are expected to expose a nameplate with rated characteristics.

%The AAS should provide a voltage property in volts in any included submodel as well as a property for the maximum rotation speed. We then retain only those whose voltage value lies within the target range. 

%Listing \ref{lst:query} presents a pseudo-SPARQL query that implements this filtering pattern and returns only AAS that contain the required submodel. The result of this step is a set of plausible AAS candidates that are then passed to the next stage.
%The query is giving back only AAS, all AAS that contain the missing instance.
%The query is giving back only AAS, where a submodel (subject) has a property (according to the predicate) named Voltage (object). The remaining lines filter the AAS further.
%The result of this pre-discovery step is a set of plausible AAS.
%These AAS are then taken to the next step. %\todo{checken ob "reasoning" der richtige begriff}

%     AAS hasSubmodel DigitalNameplate .

\begin{lstlisting}[caption=Pseudo-SPARQL Query, morekeywords={SELECT,WHERE},
                    keywordstyle=\bfseries, label={lst:query}]
SELECT AAS
WHERE {
    ?aas hasSubmodel ?sm .
    ?sm hasIdShort TimeSeriesData
}
\end{lstlisting}

%\begin{lstlisting}[caption=Pseudo-SPARQL Query, morekeywords={SELECT,WHERE},
    %                keywordstyle=\bfseries, label={lst:query}]
%SELECT AAS
%WHERE {
 %   ?submodel hasProperty Voltage .
  %  Voltage value ?voltageValue .
   % FILTER(220 <= ?voltageValue <= 240)

%    ?submodel hasProperty MaxRotationSpeed .
%}
%\end{lstlisting}

%For the sake of clarity and comprehensibility, the SPARQL query shown in this paper is not aligned with the specific structure of a real AAS implementation. It serves as a conceptual example to demonstrate how semantic pre-filtering can be based on characteristic features rather than fixed vocabulary terms.

\subsection{Vectorization via RDF2vec}

After SPARQL pre-discovery, we embed the filtered AAS graphs with RDF2vec to obtain vector representations suitable for similarity search. On the filtered graphs, we train or reuse RDF2vec embeddings to obtain vector representations of entities. RDF2vec extends the DeepWalk principle to RDF data by performing random or breadth-first graph walks over entities and relations\cite{Ristoski.2016}. The resulting sequences are then treated analogously to text sentences, see Fig. \ref{fig:rdf2vec}, and embedded using neural language models. In this way, entities occurring in similar structural contexts receive similar vector representations, enabling a quantitative comparison of AAS based on semantic structure. % Possible modeling objectives and challenges regarding RDF2vec are mentioned in Chapter \ref{heading:ModelingObjectives}.

%For AAS-level similarity, the preferred representation is the embedding of the AAS root resource, provided a stable AAS root is defined. To accommodate cases where a single root is not available or a finer-grained view is desirable, we remain optionally open to a subgraph representation, obtained by aggregating embeddings of selected nodes. This preserves modularity without committing to a specific schema.

%\subsection{Similarity measures for vectors}\label{heading:Similarity}

%The similarity between the query AAS and each candidate is computed by using the standard cosine similarity measure \cite{doi:10.3233/SW-180317}. The similarity between two entities \(e_1\) and \(e_2\) is calculated as the cosine similarity between the vectors \(V_1\) and \(V_2\)\cite{doi:10.3233/SW-180317}:

%\[sim(e_1, e_2)=\frac{V_1 \cdot V_2}{\|V_1\| \cdot \|V_2\|}\]

%Other distance measures can be substituted without changing the overall pipeline. Some works report better results when using other distance metrics, for example Euclidean distance (L1)\cite{pub15316}. It measures the straight-line distance between two points or vectors in an n-dimensional space.

%\[
%Euclidean Distance = d(X,Y) = \sqrt{\sum_{i=1}^{n} (x_i - y_i)^2}
%\]

%Other distance metrics like the Manhattan distance metric are described in \cite{deCarvalho}.

%\section{Modeling objectives} \label{heading:ModelingObjectives} % vllt Implementation Outlook oder so

\section{Challenges and Implementation} \label{heading:ModelingObjectives}
%--> Semantic differencing --> RWTH Aachen anscheinend

%For the implementation of the approach two aspects have to be analyzed. 
This chapter outlines the primary research objectives of the RDF2vec method and the process for determining the similarity threshold for the resulting AAS.

\subsection{RDF2vec method}

The approach leads to the following learning objectives and challenges for RDF2vec in knowledge graphs\cite{Van.2023}:
\begin{itemize}
    \item RDF2vec aims to preserve the topological structure of an RDF graph within a continuous vector space. By translating graph walks into linear sequences, it enables neural models to capture neighborhood relations between entities. The risk hereby is a focus on structural context while semantic similarities are neglected. For instance, two AAS using alternative property names may represent equivalent semantics, yet RDF2vec embeddings could diverge if the syntactic structures differ \cite{Van.2023}.
    \item RDF graphs contain literal values such as numbers and dates. Standard RDF2vec models primarily operate on entities and relations, often neglecting literals such as numerical values or textual attributes, which are essential in AAS contexts for describing quantitative parameters such as voltage, current, or torque. Neglecting these literals can lead to incomplete semantic representations. There has been previous work on that topic\cite{gesese2020surveyknowledgegraphembeddings, kristiadi2019incorporatingliteralsknowledgegraph}.
\end{itemize}
Addressing these challenges is one of the key learning objectives of this work, as well as the opportunities resulting from it after the implementation.

\subsection{Similarity metric and threshold}\label{heading:Similarity}
Another aspect that is to be investigated is the choice of the similarity metric for the resulting vectors.
The similarity between the query AAS and each candidate is computed by using the standard cosine similarity measure \cite{doi:10.3233/SW-180317}. The similarity between two entities \(e_1\) and \(e_2\) is calculated as the cosine similarity between the vectors \(V_1\) and \(V_2\)\cite{doi:10.3233/SW-180317}:

\begin{equation}
    \mathrm{sim}(e_1, e_2)=\frac{V_1 \cdot V_2}{\|V_1\| \cdot \|V_2\|}
\end{equation}

Other distance measures can be substituted without changing the overall pipeline. Some works report better results when using other distance metrics, for example Euclidean distance \cite{pub15316} or Manhattan distance \cite{deCarvalho}. % \todo{was bedeutet L1? Kann das weg?}
Euclidean distance measures the straight-line distance between two points or vectors in an n-dimensional space.
%Other distance metrics like the Manhattan distance metric are described in \cite{deCarvalho}. 
After defining the metric, a fitting threshold determining the number of output AAS must be defined.
%it is necessary to define a fitting threshold that determines the number of output AAS.

\begin{equation}
   \mathrm{Euclidean Distance} = d(X,Y) = \sqrt{\sum_{i=1}^{n} (x_i - y_i)^2}
\end{equation}

Let \(G_q\) be the query AAS that the user in Fig. \ref{fig:concept} wants to fill with more data and $\mathcal{G} = \{G_i\}$ the candidate set obtained after SPARQL pre-filtering in the first step of the concept.
A similarity index $s(G_q, G_i)$ as the cosine similarity between vector representations of \(G_q\) and \(G_i\) with  \(V_1\) and \(V_2\) being identical for $s=1$ is computed.
Three decision policies regarding the output of a \textit{similar AAS} are considered:
\begin{itemize}
    \item Threshold policy: Return all \(G_i\) with $s(G_q, G_i)\geq t$.
    \item Top-k policy: Return the k most similar AAS.
    \item Hybrid policy: Return $\mathcal\{G_i | s(G_q, G_i)\geq t\}$, forming set \(R_t\). If $|R_t|<k$, the set is filled up to k by decreasing s.
\end{itemize}
    
The method to compute the variables k and t are to be determined by existing studies or empirically.
Depending on the metric selection, it is to be investigated whether a mapping from the original domain to the target decision scale of $[0,1]$ is necessary, as for example the cosine similarity is defined in the interval $[-1,1]$. 

\section{Discussion}
%\hl{Since the paper is mainly focusing on the identification of the research gap and the definition of a concept solving it, this Chapter is addressing the tesing pipeline and open challanges regarding the performance, the use and alternatives of this approach.}

\subsection{Testing Pipeline}
To ensure verifiability and reproducibility, a structured testing pipeline is defined.
First, a toy dataset with a small number of AAS instances including varied vocabularies, renamed properties, and partial template implementations will be constructed. 
This dataset enables a concrete step-by-step walkthrough of the hybrid pipeline:
\begin{enumerate}
    \item Transformation of an AAS to an RDF graph,
    \item SPARQL constraint filtering,
    \item Generation of RDF2vec walks,
    \item Ranking of retrieved AAS by similarity.
\end{enumerate}

Second, the embedding model used in the experiments can either be pre-trained once on the entire AAS repository or trained specifically for the toy dataset. %to enable full reproduction. 
Embedding training is therefore not part of the evaluation itself, but part of the reproducibility setup. In practical deployments, embeddings would be trained once upfront and reused throughout subsequent testing phases.
Parameters that are essential for the reproduction pipeline are graph walk configurations, embedding parameters and similarity thresholds. 

\subsection{Weaknesses and failure modes}
The proposed approach relies on several assumptions that introduce potential weaknesses.
\subsubsection{Usability issues}
SPARQL filtering requires a level of structural knowledge about missing components of the incomplete AAS, typically a name or identifier of a submodel or its key properties. A central challenge is the high variability of modeling
practices across manufacturers. It limits applicability in scenarios where users cannot express structural constraints. Applying a filter then may preemptively exclude semantically valid AAS instances from the feasible set.

\subsubsection{Vocabulary and modeling diversity}
SPARQL filtering reduces complexity but does not eliminate the risk of irrelevant or marginally related candidates when vocabularies deviate strongly. Embedding-based similarity reduces terminological variation, but the combined structural–semantic retrieval may still fail when assets are semantically close yet modeled using substantially different structural patterns.
This may lead receiving low similarity scores for similar AAS.
Another challenge is harmonizing vocabularies such as ECLASS and IEC CDD, which will be addressed through standardization.
The development of the harmonization should be monitored and integrated in the outlook\footnote{https://github.com/eclipse-esmf/esmf-semantic-aspect-meta-model/issues/340}.
%\footnote{ https://github.com/admin-shell-io/aas-specs-metamodel/issues/422}
%\hl{Welche Fußnote???}

\subsubsection{Scalability and computational cost}

Embedding-based similarity requires generating RDF2Vec embeddings over potentially large AAS knowledge graphs. Depending on the number of AAS instances, the size of their submodels, and the graph walk parameters, the computational cost can become significant. Pre-filtering via SPARQL reduces the candidate set, but the full workflow may still face scalability constraints in industrial-scale repositories with thousands of heterogeneous AAS\cite{10825006}.
It is therefore advisable to consider and evaluate different RDF2Vec implementations\cite{Paulheim.2023}.
While the approach is expected to work for small and medium-sized repositories, the threshold at which embedding generation or similarity computation becomes inefficient has not yet been empirically determined and remains an open research question.

\section{Conclusion and Outlook}

This paper introduced a hybrid graph matching approach to enhance semantic discovery among Digital Twin representations based on the AAS. The proposed method combines rule-based pre-filtering using SPARQL with embedding-based similarity computation via RDF2vec, thus integrating symbolic reasoning and statistical learning into a unified framework. This fusion enables the comparison of AAS instances not only by structural similarity but also by semantic context, addressing limitations of purely syntactic or purely embedding-based techniques enabling re-usage of existing AAS instances in industry applications like modeling a new product.

By leveraging SPARQL, domain-specific constraints and ontology-based semantics can be incorporated directly into the matching process, ensuring that only contextually relevant candidates are considered.
RDF2vec embeddings capture the relationships between entities, allowing for context-aware similarity scoring across vocabularies and modeling conventions. 
This concept demonstrates the potential of combining rule-based and data-driven querying for Digital Twin interoperability. 
The proposed approach contributes toward reducing manual integration effort, supporting asset discovery, and enabling knowledge reuse across manufacturers and domains.

Future work will focus on several directions. First, the approach will be empirically validated using larger and more diverse AAS datasets. Second, the modeling objectives regarding the RDF2vec method will be explored to improve representation and understanding of numerical properties in AAS, as well as ensuring the capture of semantically related information. Lastly, case studies will lead to investigation of the best fitting similarity parameters.

%This work presents a hybrid graph matching approach for comparing heterogeneous AAS models by combining rule-based SPARQL pre-filtering with embedding-based similarity calculation using RDF2Vec. The expectation of this approach is to enable semantics-aware retrieval and improve the comparability of digital representations across heterogeneous AAS datasets.

%The particularly interesting part of this work is the fusion / use of information ... and the AAS specific use case...  

%In future work, the approach can be extended with advanced embedding methods beyond RDF2Vec, which claim integrate structural and latent semantic features into the embedding process. ----> \textbf{FLAG: A similar structural and semantic integrated method for RDF entity embedding}
\section*{Acknowledgements}
This Project is supported by the Federal Ministry for Economic Affairs and Energy (BMWE) on the basis of a decision by the German Bundestag -- CompAAS -- KK5311208RG4. 

Partly funded by the Federal Ministry for Economic Affairs and Energy (BMWE) -- Factory-X -- 13MX001L (\url{https://factory-x.org}) and growING -- 13IPC036G.

%-- CompAAS -- KK5311208RG4, the Federal Ministry for Economic Affairs and Energy (BMWE) -- Factory-X -- 13MX001L (\url{https://factory-x.org}), and by the Federal Ministry for Economic Affairs and Energy (BMWE) -- growING -- 13IPC036G (\url{https://www.growingdigitaltwin.de}).

%Partly funded by the Federal Ministry for Economic Affairs and Energy (BMWE) -- Factory-X -- 13MX001L. Website \url{https://factory-x.org}, the Federal Ministry for Economic Affairs and Energy (BMWE) -- growING -- 13IPC036G. Website \url{https://www.growingdigitaltwin.de}, and by the Innovation Campus Future Mobility - Website: \url{https://www.icm-bw.de/en/}.

\vspace{12pt}

\bibliographystyle{IEEEtran}
\bibliography{ac_bib}

\vspace{12pt}

\end{document}